\documentclass[pre,twocolumn,superscriptaddress, showpacs,amsmath,amssymb]{revtex4}
\usepackage{graphicx}
\usepackage{subfigure}
\begin{document}
\title{The Rabi Hamiltonian in the dispersive regime}
\author{Titus Sandu}
\affiliation{National Institute for Research and Development in Microtechnologies-IMT,
126A, Erou Iancu Nicolae Street, 077190, Bucharest, ROMANIA}
\email{titus.sandu@imt.ro}
\date{\today}
\begin{abstract}
The Rabi Hamiltonian is studied in the dispersive regime and ultra-strong coupling. We employ a recent unitary transformation to 
obtain not only the approximate Hamiltonian and its energy levels but also its eigenfunctions. The relationship of the 
approximation with other regimes and their approximations are also discussed.
\end{abstract}
\pacs{42.50.Md, 42.50.Hz, 63.20.Kr, 85.25.Cp, 85.85.+j}
\maketitle 
\section{Introduction}
The Rabi Hamiltonian was used initially to describe the 
interaction of nuclear spins with magnetic fields \cite{Rabi1937}, but it is also used to model the ammonia molecule \cite{Bersuker1989} or 
electrons coupled with a phonon mode in a crystal lattice \cite{Holstein1959}. More recently, beside the interaction of atoms with an 
electromagnetic field in a cavity \cite{Scully1996} the Rabi Hamiltonian describes the interaction of superconducting qubits 
with a nanomechanical resonator \cite{Schwab2005}, with a transmission line resonator 
\cite{Wallraff2004}, or an LC resonator \cite{Chiorescu2004}. 
The Rabi Hamiltonian expresses the interaction of a two-level system (TLS) with a single boson mode and has the 
following form:
\begin{equation}
\label{eq1}
H = \omega {a^\dag a} + \Omega \sigma _z+ \lambda ( {a^\dag + a})( {\sigma_{-} + \sigma_{+}}),
\end{equation}
\noindent
where $\hbar = 1$, $\sigma _{x}$, $\sigma _{y }$, and $\sigma _{z }$ are spin $1/2$ 
matrices, $\sigma _{\pm}=\sigma _{x} \pm i\sigma _{y}$, $a^\dag $ and $a$ are the creation and annihilation operators of the quantum 
oscillator. 

The apparent simplicity of (\ref{eq1}) led to the conjecture 
about its solvability  \cite{Reik1986} that only recently has been fully proved  \cite{Braak2011}. Despite the fact that the Rabi Hamiltonian 
is completely solvable, its general solutions cannot be easily grasped in simple 
terms. In quantum optics with atoms in a cavity also known as cavity quantum electrodynamics (cavity QED) one encounters a 
regime of quasi-resonance ($\omega \approx \Omega$) and a  
coupling strength $\lambda/\omega$ between $10^{-7}$ and $10^{-5}$. This regime is very well described by the Jaynes-Cummings model 
which is simpler and solvable  \cite{Jaynes1963}.
The Jaynes-Cummings model uses the rotating wave 
approximation (RWA) in (\ref{eq1}), such that the counter-rotating wave term 
$a^\dag \sigma_{+} + a\sigma _{-}  $ is completely 
ignored \cite{Scully1996}. 
Moreover, in the circuit QED \cite{Schwab2005,Wallraff2004,Chiorescu2004,Blais2004}, where the TLS 
is associated with a qubit, one encounters a dispersive regime (i. e., large detuning  $|\omega - \Omega| \sim \omega + \Omega $) 
with a ultra-strong coupling 
$\lambda/\omega \sim 0.1$, where the Jaynes-Cummings model fails \cite{Niemczyk2010,Forn-Diaz2010,Fedorov2010}. 
A more elaborate approximation based on the polaron 
transformation \cite{Holstein1959,Sandu2009} is the generalized RWA, which works well 
in the region of zero and large detuning and ultra-strong coupling \cite{Irish2007}. 

Recently 
Zueco {\it{et al.}} \cite{Zueco2009} used the same technique of the unitary transformations to include the counter-rotating wave term 
with an improved accuracy of energy levels for a wider range of parameters. 
We build on that work and in Section 2 we provide 
not only the approximate Hamiltonian and its eigenvalues but also its eigenfunctions to the second order in the coupling constant. 
In section 3 we discuss comparatively the regimes of slow and fast TLS with respect to the oscillator and in Section 4 we conclude the work.

\section{The approximate Hamiltonian in the dispersive regime}
Quite often a general Hamiltonian $H$ can be decomposed as $H = H_0 + \lambda 
\,W$, where $H_0 $ is diagonalizable and $\lambda$ is a small real parameter. 
The ultimate goal of the unitary transformation 
approach is to find an anti-hermitian operator $S$ such that the unitary 
transformation $U = e^{\lambda S}$ diagonalizes $H$.  
However, the complete diagonalization is not always possible but it can be done 
successively beginning with the first order of $\lambda$. The requirement to eliminate $\lambda W$ in the first order 
leads to the form of 
$S$ that is given by $S = - i\, \lambda \mathop {\lim }\limits_{\varepsilon \to 0_ + } \int\limits_{ - 
\infty }^0 {e^{\varepsilon t}W_I \left( t \right)dt} $, 
where $W_I \left( t \right) = e^{iH_0 t}We^{ - iH_0 t}$ \cite{Sigmund1974}. If we choose 
$H_0 = \omega a^\dag a + \Delta \sigma _z $ and $\lambda \,W = \lambda \left( {a 
+ a^\dag } \right)\left( {\sigma _ - + \sigma _ + } \right)$ we can calculate $S$ as being
\begin{equation}
\label{eq6}
S =  - {\frac{2 i \omega}{\Delta^2 - \omega^2 }} 
\left( {\frac{a - a^\dag}{i} } \right)\sigma _x - 
{\frac{2 i \Delta}{\Delta^2 - \omega^2 }} \left( {a + 
a^\dag } \right)\sigma _y .
\end{equation}
The use of (\ref{eq6}) in the transformation $U_1 = e^{\lambda S}$ changes the Rabi Hamiltonian into   
\begin{equation}
\label{eq7}
\tilde{H} = \omega a^\dag a + \Delta \sigma _z + \frac{2\lambda ^2\Delta 
}{\Delta ^2 - \omega ^2}\left( {a + a^\dag } \right)^2\sigma _z + \frac{\lambda ^2\omega 
}{\Delta ^2 - \omega ^2} + O\left( 
{\lambda ^3} \right).
\end{equation}
We notice that the small parameter is 
$\tilde{\lambda}=\lambda^2/(\Delta ^2 - \omega ^2)$ and if $\tilde{\lambda}  \ll 1$ we can retain just the first 
three terms of (\ref{eq7}). Then we expand $\left( {a + a^\dag } \right)^2$, make 
some arrangements and remove completely the part of Hamiltonian proportional with ${a^2 + {a^\dag}^2}$. This 
is possible with the 
unitary transformation $U_2 = e^{\hat{\beta} \sigma_{z}({a^2 - {a^\dag}^2})}$, where $ \hat{\beta} = \begin{pmatrix}
  \beta_{+} & 0 \\
  0       &\beta_{-}
 \end{pmatrix}$ and $ tanh(2\beta_{\pm}) = \frac{\pm 2\tilde{\lambda}\Delta }{\omega \pm 2\tilde{\lambda}\Delta}$. 
Thus, the approximate form  of (\ref{eq7}) is
\begin{widetext}
\begin{equation}
\label{eq10}
H_{1} = a^\dag a\sqrt {\omega ^2 + {8\tilde{\lambda}\omega \Delta }\sigma _z } + \frac{\sqrt {\omega ^2 + {8\tilde{\lambda}\omega \Delta }\sigma _z }-\omega}{2} + (1 + 2\tilde{\lambda})\Delta\sigma_z + \tilde{\lambda}\omega. 
\end{equation}
\end{widetext}
To be fully consistent with the 
approximations made so far we expand the radical in Eq. (\ref{eq10}) and keep just the first two terms in the series. Thus, the oscillator 
frequencies are $\omega_{\pm} = \omega \pm 2\tilde{\lambda} \Delta $ and the Hamiltonian (\ref{eq10}) turns into the Hamiltonian  of the 
dispersive regime and ultra-strong coupling
\begin{equation}
\label{eq11}
H_{approx} = a^\dag a \begin{pmatrix}
  \omega_{+} & 0 \\
  0       &\omega_{-}
 \end{pmatrix} + (1 + 4\tilde{\lambda})\Delta\sigma_z + \tilde{\lambda}\omega. 
\end{equation}
It is easy to see that the eigenvalues of (\ref{eq11}) are
\begin{equation}
\label{eq12}
E_{\pm n}=n\omega_{\pm} \pm \frac{1 + 4\tilde{\lambda}}{2}\Delta + \tilde{\lambda}\omega.
\end{equation} 
We also calculate the eigenvectors of (\ref{eq11}) in the original Schr\"odinger picture. In order to do so 
we need to evaluate the action of $ e^{\lambda S}$ on any vector in the Hilbert space of the problem. The action of 
$ e^{\lambda S}$ is rather cumbersome but a meaningful expression can be obtained with the Zassenhaus formula \cite{Magnus1954}, which, for 
any operator $A$ and $B$ reads $e^{\lambda(A + B)} = e^{\lambda A}e^{\lambda B}e^{-\frac{\lambda^2}{2}[A,B]} \ldots \ldots$\,\,.
Thus, the eigenvectors of (\ref{eq11}) are    
\begin{widetext}
\begin{equation}
\label{eq14}
|\Psi_{\pm n}> = {e^{-i\frac{2\lambda \omega}{\Delta^2 - \omega^2}\frac{a - a^{\dag}}{i}\sigma_x} 
e^{-i\frac{2\lambda \Delta}{\Delta^2 - \omega^2} ({a + a^{\dag}})\sigma_y} 
e^{(\frac{2\lambda^2 \omega \Delta }{(\Delta^2 - \omega^2)^2}+\hat{\beta})(a^2 - {a^{\dag}}^2) \sigma_z} }|n(\omega_{\pm})> s_{\pm}.
\end{equation}
\end{widetext}
In Eqs. (\ref{eq12}) and (\ref{eq14}) $n$ is a natural number, $|n(\omega_{\pm})>$ are the $n$th eigenvectors of the quantum oscillator with frequencies 
$\omega_{\pm}$, and $s_{\pm}$ are the eigenvectors of $\sigma_z$, i. e., $\sigma_zs_{\pm}=(\pm1/2)s_{\pm}$.

\section{Discussions}
The energy levels given by Eq. (\ref{eq12}) have been compared with the exact eigenvalues of (\ref{eq1}) in the paper of Zueco {\it{et al.}} 
\cite{Zueco2009}, 
where it has been proved a very good match between (\ref{eq12}) and the eigenvalues of (\ref{eq1}) for a wide range of parameters in the dispersive regime. 
On the other hand, the accuracy of eigenfunctions has been checked in Ref. \cite{Agarwal2013} by calculating their fidelity with respect to the exact 
numerical calculated eigenfunctions. 
The comparison of the eigenfunctions have been performed in two limiting cases of the dispersive regime: slow TLS and fast oscillator 
($\Delta \ll \omega $) and slow oscillator and fast TLS ($\Delta \gg \omega $).

In the slow TLS and fast oscillator regime ($\Delta \ll \omega $) Agarwal {\it{et al.}}\cite{Agarwal2013} have found that the 
eigenvectors of Hamiltonian (\ref{eq11}) in Schr\"odinger picture have a poor fidelity with respect to the exact eigenvectors. Moreover, 
their calculations indicated that the fidelity of the eigenvectors provided by the adiabatic approximation 
\cite{Irish2007,Sandu2009} is excellent. We compare our results given by (\ref{eq14}) with the adiabatic eigenvectors.
We recall that the adiabatic approximation is obtained using the polaron transformation 
$e^{2\frac{\lambda}{\omega}(a - a^{\dag})\sigma_x}$ and neglecting the spin non-diagonal term \cite{Sandu2009}. 
Hence, the adiabatic Hamiltonian and its eigenvectors are, respectively, 
\begin{equation}
\label{eq17}
H_{adiab} = {\omega} a^\dag a + \Delta \sigma _z cos(\frac{2\lambda}{\omega}\frac{a - a^{\dag}}{i}) - \frac{\lambda^2}{\omega},
\end{equation}
\begin{equation}
\label{eq18}
|\Psi_{{\pm n}\_adiab}> = {e^{2\frac{\lambda}{\omega}(a - a^{\dag})\sigma_x}}|n(\omega)>s_{\pm}.
\end{equation}
When $\Delta \ll \omega $ both $\lambda/\omega$ and $\Delta/\omega$ are small, therefore (\ref{eq14}) becomes
\begin{widetext}
\begin{equation}
\label{eq19}
|\Psi_{\pm n}> = {e^{ \frac{2\lambda}{\omega}({a - a^{\dag}})\sigma_x} 
e^{2i\frac{\lambda \Delta}{\omega^2}({a + a^{\dag}})\sigma_y} 
e^{\frac{\lambda^2 \Delta }{\omega^3}(a^2 - {a^{\dag}}^2)(2 \sigma_z+1)}}|n(\omega_{\pm})>s_{\pm}
\end{equation}
\end{widetext}
In Eq. (\ref{eq19}) the part that contains the squeeze operator deviates from the unit operator by an amount of the same order 
as the overlap $<n(\omega)|n(\omega_\pm)>$ differs from unity, i. e., $O(\frac{\lambda^2 \Delta }{\omega^3})$, thus their deviation 
from 1 can be safely discarded. It is easy now to evaluate the fidelity between (\ref{eq18}) and (\ref{eq19})
as $f = |<\Psi_{\pm n\_adiab}|\Psi_{\pm n}>|^2= 1-O(\frac{\lambda \Delta}{\omega^2})$,
which is very close to 1. It would imply that the fidelity 
of (\ref{eq14}) is close to one and in stark contrast with Ref. \cite{Agarwal2013}. One possible explanation of 
this discrepancy might be the way $e^{\lambda S}$ is calculated to obtain the eigenfunctions. We adopted an exponential approximation based on the Zassenhaus formula, 
but if one adopts a series expansion of $e^{\lambda S}$, the fidelity would be $f = 1 - O(\lambda/\omega)$, which is sensibly below unity and consistent with the conclusions of 
Agarwal {\it{et al.}}
  
In the region $\Delta \gg \omega $ (fast TLS and slow oscillator) one can also invoke an adiabatic approximation as long as the TLS and the oscillator 
are on two different 
energy scales. This situation is more often encountered in molecules and crystals \cite{Bersuker1989}.  
The new adiabatic regime can be obtained by a unitary transformation that also works well for the regime of deep ultra-strong coupling when 
$\lambda \ge \omega$ \cite{Sandu2006}. This adiabatic approximation generates an adiabatic potential with two sheets. The lower sheet has 
two minima if $\lambda \ge \sqrt{\omega \Delta}$ and only one minimum if $\lambda < \sqrt{\omega \Delta}$ \cite{Sandu2006}.  
When $\lambda \ge \sqrt{\omega \Delta}$ an approximate solution based on the displaced oscillators generated by the the adiabatic potential is given in 
Ref. \cite{Irish2014}. 
However, the case of the adiabatic potential with just one minimum can be described by the Hamiltonian (\ref{eq11}) 
of which both the eigenvalues \cite{Zueco2009} and the eigenvectors \cite{Agarwal2013} are well reproduced by Eqs. (\ref{eq11})-(\ref{eq14}). We note here that in the 
regime $\Delta \gg \omega $ a series expansion of $e^{\lambda S}$ gives rather similar results with Eq. (\ref{eq14}). Moreover, one can 
easily check that the curvatures of the two adiabatic sheets in the origins are just ${\omega_-}^2$ and ${\omega_+}^2$, respectively. The eigenvector 
(\ref{eq14}) exhibits also a certain degree of squeezing. In fact, the largest squeezing is encountered at 
$\lambda = \lambda_c= \sqrt{\omega \Delta}$ \cite{Sandu2003}, where the system undergoes a sharp transition\cite{Ashhab2013}.

\section{Conclusions}
In this work we studied the Rabi Hamiltonian in the dispersive regime and ultra-strong coupling. It was used a unitary transformation that 
takes into account terms beyond the rotating wave approximation. We are able to build the approximate Hamiltonian with its energy levels and 
its eigenfunctions. It turns out that an exponential approximation of the eigenfunctions is better suited than the approximation made by series expansion. We 
also compare and discuss this 
approximation with respect to other approximations in the regimes of fast and slow two-level system.

\begin{acknowledgments}
This work was supported by a grant of the Romanian National Authority for 
Scientific Research, CNCS -- UEFISCDI, project number PNII-ID-PCCE-2011 
-2-0069. 
\end{acknowledgments}

% Create the reference section using BibTeX:
%\bibliography{bib_article-good}
%\bibliography{./T_Sandu}
%%\begin{thebibliography}{4}
%%\end{thebibliography}

\end{document}